\documentclass{mn2e}
\usepackage{psfig}
\usepackage{mnras_cite}
\newcommand{\simgt}{\lower.5ex\hbox{$\; \buildrel > \over \sim \;$}}
\newcommand{\simlt}{\lower.5ex\hbox{$\; \buildrel < \over \sim \;$}}
\begin{document}

\title[Isocurvature Fluctuations \& Early Reionisation]
{Isocurvature Fluctuations Induce Early Star Formation}
\author[Sugiyama, Zaroubi, Silk]{Naoshi Sugiyama $^1$, Saleem Zaroubi$^2$, Joseph
Silk$^3$\\
$^1$Division of Theoretical Astrophysics, National Astronomical Observatory Japan, Mitaka, Tokyo
181-8588,Japan \\
$^2$Max Planck Institute for Astrophysics, Karl Schwarzschild Str.1,
D-85741, Garching, Germany \\
$^3$Astrophysics, University of Oxford, Keble Road, Oxford OX1 3RH}

\maketitle

\begin{abstract}
The early reionisation of the Universe inferred from  the WMAP
polarisation results, if confirmed,  poses a problem 
for  the hypothesis that  scale-invariant
adiabatic density fluctuations account for large-scale structure and
galaxy formation. One can only generate the required amount
of  early star formation if extreme assumptions are made about the
efficiency  and nature of  early reionisation.  We develop 
an alternative hypothesis that invokes an additional component 
 of  a non-scale-free isocurvature power spectrum 
together with the scale-free adiabatic power spectrum for
inflation-motivated primordial density fluctuations. Such a
component  is constrained by the Lyman alpha forest observations, can
account for the small-scale power  required by spectroscopic
gravitational lensing, and yields  a source of early star formation
that can reionise the universe at $z\sim 20$ yet becomes an
inefficient source of ionizing photons
by $z\sim 10, $ thereby allowing  the conventional adiabatic fluctuation
component to reproduce the late thermal history of the intergalactic medium.
\end{abstract}

\begin{keywords}
cosmology: theory -- early universe -- large-scale structure of Univerese --
intergalactic medium -- galaxies: formation -- stars: formation 

\end{keywords}

\section{Introduction}

There have been major recent advances in cosmology with impact on galaxy
formation theory. These include the detection of the
temperature--polarisation cross power spectrum for the CMB by 
WMAP~\cite{kogut},
and measurements both of the CMB temperature power spectrum and the
underlying matter power spectrum with unprecedented accuracy, utilizing
the WMAP, 2DF and quasar Ly$\alpha$ absorption line data
sets~\cite{bennett,spergel}.  Two results that have received
considerable attention are the optical depth of the universe
$\tau=0.17\pm 0.03,$ which requires that the epoch of reionisation occurs
at z=15-20 from WMAP~\cite{kogut,spergel}, and the rolling spectral
index $\mathrm{d}n/\mathrm{d}lnk = -0.03 \pm 0.01$
for approximately scale-invariant
density fluctuations for a combination of WMAP, 2DF and Ly$\alpha$
data~\cite{spergel}.

There is some tension between these results: if both are correct, it is
difficult to understand how recombination occurred so early without some
modification of the canonical model of primordial, nearly
scale-invariant Gaussian adiabatic density
fluctuations~\cite{ciardi,fukugita,haiman,som2}. 
In fact, a new
Ly$\alpha$ absorption data set from the SDSS quasars has independently
found evidence for a rolling spectral index~\cite{seljak}, although an
independent analysis of the same data does not reproduce sufficiently
small error bars to confirm this result~\cite{abazajian}.

The Ly$\alpha$ lines measure power in the underlying matter power spectrum 
on a comoving scale of around 1 Mpc. The results are however subject to 
bias, since one has to be confident that the gas is relatively unperturbed 
by feedback, such as is seen in the vicinity of Lyman break galaxies to 
Mpc distances. Hence it is of particular interest to consider another 
measure  of the power spectrum on even smaller comoving scales, $ 10^{-2} $
to 0.1 Mpc. This comes from spectroscopic
 gravitational lensing of quasar emission line region on several
 scales, the  magnification ratios requiring  and constraining 
substructure in  the  
massive lensing halos~\cite{met}.
One needs substantial power in objects of  $10^6$ to $10^9 \rm M_\odot$, 
amounting to between  4 and 7  percent of the 
galaxy surface density, and this cannot easily be accommodated in
the usual CDM models with standard elliptical  isothermal lens mass
profiles.
Previous estimates of halo substructure from gravitational lensing 
using simple lens models are highly uncertain~\cite{dal}.
 Moreover the numerical simulations of halos suggest that, for nearly
 scale-invariant initial conditions, small-scale power may largely be
 erased in the inner halos. One is in a dangerous regime of the spectrum
 where $n_{\rm eff} \approx -3, $ and tidal destruction is
 effective. Indeed it is not clear whether the simulations have enough
 resolution to adequately address the question of the survival of
 small-scale substructure.

We propose an alternative prescription for small-scale power that 
satisfies all observed constraints and unambiguously predicts the survival
of small-scale power, with some inflationary motivation.
We postulate that as is common to multifield inflationary  models,
both isocurvature and adiabatic fluctuation modes are generated. 
A recent model  motivated by inflation is a so-called curvaton 
model~\cite{lyth,morotaka,enq}  in which 
an additional scalar field besides the inflaton, -- 
the curvaton--,  produces fluctuations during the reheating epoch.  
If the isocurvature fluctuations were generated by 
inflaton and curvaton generated adiabatic mode, there may be 
a possibility of  having a non-scale-free isocurvature power spectrum 
together with the $n=1$ scale-free adiabatic power spectrum~\cite{moroi} .
As well as this curvaton hypothesis, 
a sub-dominant contribution of cosmic strings
induced by brane inflation in superstring theory\cite{jones}
may provide additional small-scale power.

The isocurvature contribution is described by two parameters: the
amplitude normalisation and the spectral index, which we take to be
freely assignable, but chosen to give more small-scale power than the
rolling or nearly scale invariant index adiabatic fluctuations measured
on larger scales.  We use Ly$\alpha$ forest data to constrain these
parameters.
The amount of excess small-scale power can be tuned by adjusting the 
spectral index within the allowed constraints. We normalise at 1 Mpc,
the central point of the Ly$\alpha$ probes.
Small-scale power survives because the fluctuations become nonlinear 
earlier than in the pure adiabatic case, due to the isocurvature component 
boost. The  first nonlinear fluctuations form earlier and hence lead to 
denser substructures that are resistant to tidal disruption within massive 
halos.

The two-component model has two advantages. It results in early star
formation, regardless of the spectral index measured for the adiabatic
component on large scales. Hence early reionisation can be achieved.
It also preserves small-scale power as hierarchical clustering
develops, in the form of dense $10^6 \rm M_\odot $ clumps in massive
dark halos. This helps to explain quadruple quasar lensing flux ratios
as well as bending of radio minijets.

In the remainder of this paper, we give  constraints from the Ly$\alpha$ 
forest  on the isocurvature component and we discuss the observational 
implications.  We give several applications: in addition to the
implications for the epoch of 
reionisation and  
halo microlensing, we discuss the possible 
implications for patchy   reionisation and SZ  signatures of 
very early star formation via  baryon trapping in dense early substructures,
and the clustering of early forming substructures and 
implications for the formation of the first stars.

\section{Models}

   From WMAP results, it is known that the nature of fluctuations is 
consistent with the adiabatic mode and the contribution of isocurvature
perturbations cannot be dominant on the relevant  measurement scales which are 
$k \simlt 0.05 \rm Mpc^{-1}$~\cite{spergel,bennett}.  
   
On smaller scales, $k \sim 1 \rm Mpc^{-1}$, 
the amplitude of isocurvature fluctuations is 
constrained by Ly$\alpha$ forest~\cite{crofta,nusser}.
The joint analysis of WMAP, 2dF and Ly$\alpha$ shows 
the power spectrum obtained by Ly$\alpha$ 
is significantly lower than 
the one extrapolated from a single power law $\Lambda$ CDM model
consistent with WMAP data alone~\cite{spergel}.  

Several new analyses of Ly$\alpha$ power spectrum based on the Sloan Digital
Sky Survey data are becoming available~\cite{seljak,abazajian}.  Although they
employed the same data sets, each group has obtained different values.  Among
them, Seljak (2003) provides the lowest amplitude of the power
spectrum and the smallest error bars.  Therefore here we take his value
as a reference.  We set the normalisation of the isocurvature fluctuations
to be $10\%$ of Seljak's power spectrum.  We should notice that this is
merely an  upper limit on the possible isocurvature contributions.

On the other hand, Abazajian \& Dodelson (2003)
give a relatively high central value with larger error bars.  
The central value of their power spectrum indeed exceeds the 
extrapolation from the WMAP power law $\Lambda$ CDM power spectrum.  
Here we have one extreme model assuming the difference between 
the Abazajian \& Dodelson power spectrum and 
the WMAP power law spectrum  is due to the existence of 
the isocurvature contributions.  Then we immediately obtain 
the amplitude and the power law index (which is $n=-1.7$) 
of the isocurvature power spectrum.
The corresponding power spectra are shown in Figure 1.
\begin{figure}
\centerline{\psfig{file=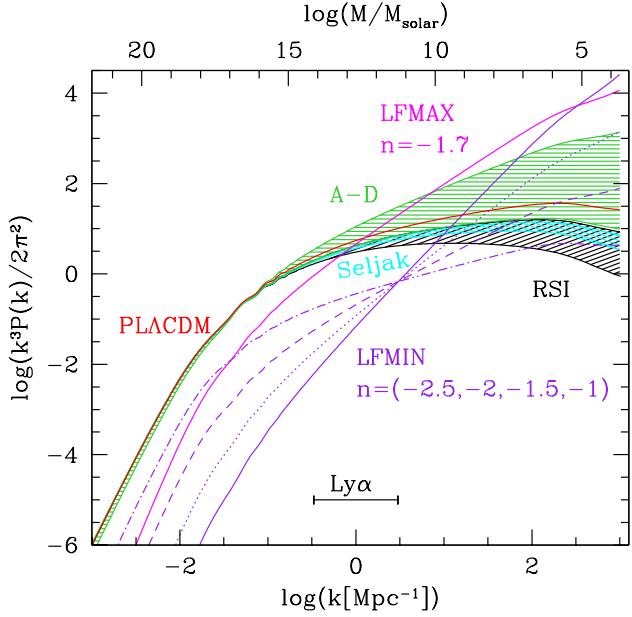,width=90mm}}
\caption{
The matter power spectra.  The red line is the WMAP best 
fitted power law $\Lambda$ CDM model (PL$\Lambda$CDM) (Spergel et al. 2003), 
the black hatched region is the running spectral index (RSI) model with 
errors fitted by WMAP, 2dF and Ly$\alpha$ (Spergel et al. 2003), and the blue and green
 hatched regions are the RSI model with the SDSS Ly$\alpha$ power spectrum 
analyzed by Seljak (2003), and Abazajian \& Dodelson (2003),
 respectively. Purple lines, which are labeled as LFMIN (Ly$\alpha$
 Forest Minimum), 
 are isocurvature power spectra with $n=-2.5, -2, -1.5$,
and $-1$ normalized to $10\%$ of Seljak's Ly$\alpha$ analysis. 
The pink line labeled as LFMAX (Ly$\alpha$ Forest Maximum) 
is the isocurvature power spectrum by assuming the excess of the central value of
Abazajian \& Dodelson (2003) is caused by isocurvature contributions.  
}
\label{fig1}
\end{figure}

\section{Reionisation}

We estimate 
the number of photons emitted from Pop III stars, which is 
a crucial element for  the reionisation process, 
following Somerville \& Livio (2003) and 
Somerville, Bullock \& Livio (2003).     

First, employing the Press-Schechter prescription, we calculate 
the fraction of the total mass in collapsed halos $F_{h}$ 
with masses greater than $M_{crit}$ and lower than $M_{vir}$.  
Here we adopt $M_{cirt}= 1\times 10^6 h^{-1}M_\odot$ and
$M_{vir}=M(T_{vir}=10^4\rm K)$.  Objects whose virial temperature
are larger than $10^4 \rm K$ can cool via atomic hydrogen and we assume
them as Pop II stars.

   From $F_h$, we can calculate the global star formation rate density 
as 
\begin{equation}
\dot{\rho_*} = e_* \rho_B 
{dF_{h} \over dt}\left(M_{vir}> M>M_{crit}\right), 
\end{equation}
where  $e_*$ is the star formation efficiency which we take 
$0.002$ for Pop III stars with $200 M_\odot$ and 
$0.001$ with $100 M_\odot$, and $\rho_B$  is the comoving background 
baryon density~\cite{yoshida}.  

Let us assume Pop III stars produce $dN_\gamma/dt = 1.6\times 10^{48} $
photons $s^{-1} \rm M_\odot^{-1}$ for a lifetime $\Delta t = 3 \times 10^6$
years~\cite{bromm1}.  Therefore we can write $dN_\gamma/dt =
N_{\gamma,0}\Theta(t)$, where $N_{\gamma,0} = 1.6\times 10^{48} $
photons $s^{-1} \rm M_\odot^{-1}$ and $\Theta(t)$ is a step function as
$\Theta(t) =1 $ for $t < \Delta t$ and $\Theta(t) = 0 $ for $t > \Delta
t$.  Using this expression, the total production rate of ionizing
photons per cubic Mpc at time $t'$ becomes
\begin{eqnarray} 
{dn_\gamma \over dt}(t') & = & \int_0^{t'}
{d N_\gamma \over dt}(t'-t)\dot{\rho_*}(t) dt  \nonumber\\
& = & 
e_* \rho_B N_{\gamma,0} \int_0^{t'}
\Theta(t'-t){dF_{h} \over dt}(t)dt \nonumber\\
& = & 
e_* \rho_B N_{\gamma,0} \int_{t'-\Delta t}^{t'}
{dF_{h} \over dt}(t)dt \nonumber\\
& = &
e_* \rho_B N_{\gamma,0} \left(F_{h}(t') -F_{h}(t'-\Delta t )\right).
\end{eqnarray}
Then we can obtain the cumulative number of 
photons per H atom as 
\begin{eqnarray} 
{n_\gamma^{cumul} \over n_H} (t_0) \hspace{-10pt}&=& \hspace{-10pt} 
{\mu m_p\over \rho_B} \int_0^{t_0} {dn_\gamma \over dt}(t') dt' \nonumber\\
&=& \hspace{-10pt} 
\mu m_p e_* N_{\gamma,0} 
\int_0^{t_0} 
\left(F_{h}(t') -F_{h}(t'-\Delta t )\right) dt' , 
\end{eqnarray}
where $n_H$ is the Hydrogen number density and $m_p$ is the
proton mass. 
Here we may employ the approximation 
$F_{h}(t') -F_{h}(t'-\Delta t) \simeq dF_{h}/dt'(t')\Delta t $, 
and we obtain 
$
{n_\gamma^{cumul} / n_H} (t_0)  \simeq 
\mu m_p e_* N_{\gamma,0} F_{h}(t_0)\Delta t. 
$
We checked this approximation works almost perfectly well for $z< 40$
when the Hubble time is longer than $\Delta t$.  

For calculating the number of photons emitted from Pop II stars,
we replace $M_{vir} > M > M_{crit}$ of equation (1) with 
$M > M_{vir}$ and set $e_* = 0.1$ and 
$dN_\gamma/dt = 8.9\times 10^{46} $ photons $s^{-1} \rm M_\odot^{-1}$ 
for a lifetime $\Delta t = 3 \times 10^6$
years~\cite{som1}.

\section{Results}

\begin{figure}
\centerline{\psfig{file=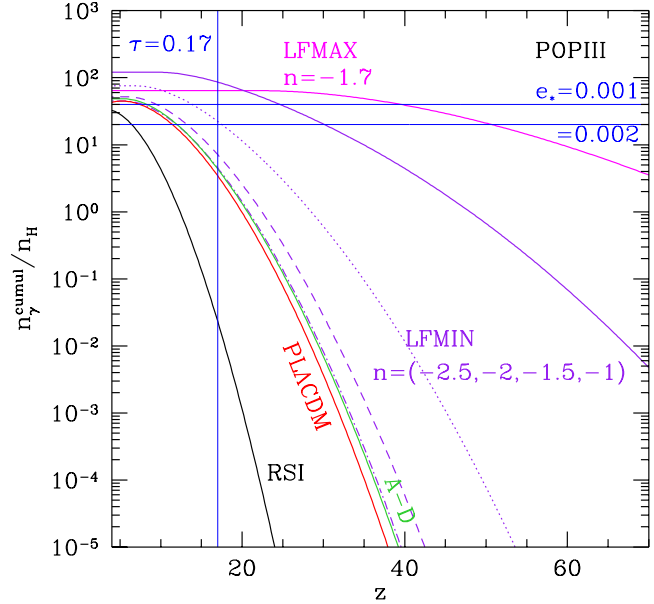,width=90mm}}
\caption{Cumulative photons emitted from Pop III stars.
Models are same as Figure 1 while we take the central value for the running spectral index model
(RSI) and Abazajian, \& Dodelson SDSS analysis (A-D).  The difference
 between power law $\Lambda$CDM (PL$\Lambda$CDM) and A-D is very small.
}
\label{fig2}
\end{figure}

In Figure 2, the cumulative number of photons emitted from 
Pop III stars
per hydrogen atom 
as a function of redshift is shown
for 
the power law $\Lambda$CDM model with WMAP parameters, 
the running spectrum index model, isocurvature models with 
the power law index $n=-2.5,-2,-1,-1.5,-1$ whose 
amplitudes are set as $10\%$ of Seljak's analysis of Ly$\alpha$
forest, the isocurvature model  with $n=-1.7$ which explains the excess of 
the power spectrum obtained by the 
analysis of Ly$\alpha$ clouds by Abazajian \& Dodelson 
against  power law $\Lambda$CDM and the model fitted with Abazajian \&
Dodelson.  
Here, we take $e_* = 0.002$.  However we can simply renormalise 
the value of y-axis for different values of $e_*$.  

\begin{figure}
\centerline{\psfig{file=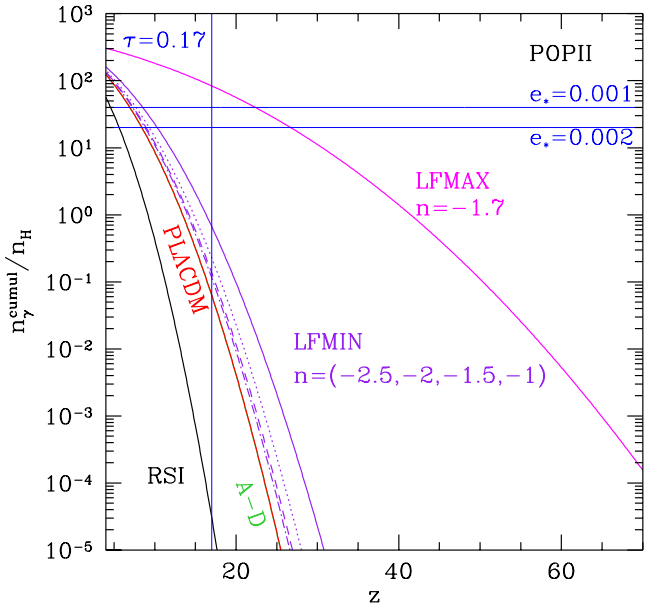,width=90mm}}
\caption{Cumulative photons emitted from Pop III stars. 
Models are same as Figure 2.  
}
\label{fig3}
\end{figure}

In Figure 3, the cumulative number of photons emitted from 
Pop II stars
per hydrogen atom 
as a function of redshift is shown.  
It is clear from Figures 2 and 3 that contributions from 
Pop II stars on early reionisation are almost negligible.

To get the ionisation fraction, we have to multiply 
$f_{esc}f_{ion}/C_{clump}$ where $f_{esc}$ is the escape fraction of 
photons from the galaxy, $f_{ion}$ is  the number of ionisations per 
UV photon, and $C_{clump}$ is the clumping factor of IGM,
respectively.  It is known that about $5$ to $20$ photons per H atom
are  required to achieve a volume-weighted ionisation fraction of 
$99\%$~\cite{haiman01,sokasian,sok1}.  We take $20$ as our reference.  
In Figure 2, this number is plotted as the horizontal line.
If we assume $e_*=0.001$ instead $0.002$, $20$ photons per H atom 
correspond to $40$ photons in this figure which we also plotted 
as a horizontal line.  The crossing of  each line with 
this horizontal line gives the epoch of reionisation. 
The vertical line in this figure is 
$z=17$ which is the reionisation epoch for WMAP with 
instantaneous reionisation. 

In Table 1, we summarize the ``reionisation'' epoch of each model.
These numbers correspond to $e_*=0.002$.  
If we employ $e_*=0.001$ and assume the cumulative number of photons
needed for reionisation is between 10 and 40 per H atom, the corresponding 
redshifts at which this occurs are given in  this Table.  
It is clear that the running spectrum index model cannot plausibly have
early enough reionisation and that power law $\Lambda$CDM is marginally 
consistent with WMAP results.  These results are consistent with 
previous works~\cite{yoshida1,ciardi,fukugita,haiman,som1}
It should be noticed that these redshifts for models with isocurvature
fluctuations besides the  A-D model are upper limits since we choose the
amplitudes of power spectra to be as large as possible without
violating 
Ly$\alpha$ constraints.

 

\begin{table}
\footnotesize
\caption{
Redshifts at which the  cumulative number of photons becomes $10, 20$ and $40$.  
Redshifts when the cumulative number of photons from PopII  becomes 
equal to the one from PopIII are also shown.  
 }
\begin{tabular}{lrrrc}
\hline
Model & 
\multicolumn{3}{c}{ $n_\gamma^{cumul} / n_H$} & 
$ \rm PopII= PopIII $\\
 &10  & 20 & 40 & \\
\hline
RSI & 8.7 & 7.3 & 5.9 & 5.2\\
PL$\Lambda$CDM & 14 & 12 & 9.6 & 6.8 \\
A-D & 15 & 12 & 10 & 6.6 \\
iso $n=-2.5$ & 15 & 13 & 10 & 7.1 \\
iso $n=-2$ & 16 & 14 & 11 & 6.8 \\
iso $n=-1.5$ & 21 & 18 & 14 & 6.0\\
iso $n=-1$ & 35 & 30 & 25 & 5.6 \\
iso $n=-1.7$ (A-D) & 59 & 51 & 40 & 19 \\
\hline
\end{tabular}
\end{table}

\section{Discussion}

We can certainly have early enough reionisation if we introduce the
isocurvature mode on small scales.  One interesting aspect is the number
of cumulative photons per H atom by PopIII stars asymptotes to a
constant at a later epoch for some models.  To illustrate this effect,
we plot the number of cumulative photons for the $n=-1$ model in Figure
4.  Here we renormalise the power spectrum to have an appropriate
reionisation epoch.  We have the number of photons about $20$ at $z=17 $
for assuming $1\%$ amplitude of Seljak's Ly$\alpha$ power spectrum and
$10$ for $1\%$.
\begin{figure}
\centerline{\psfig{file=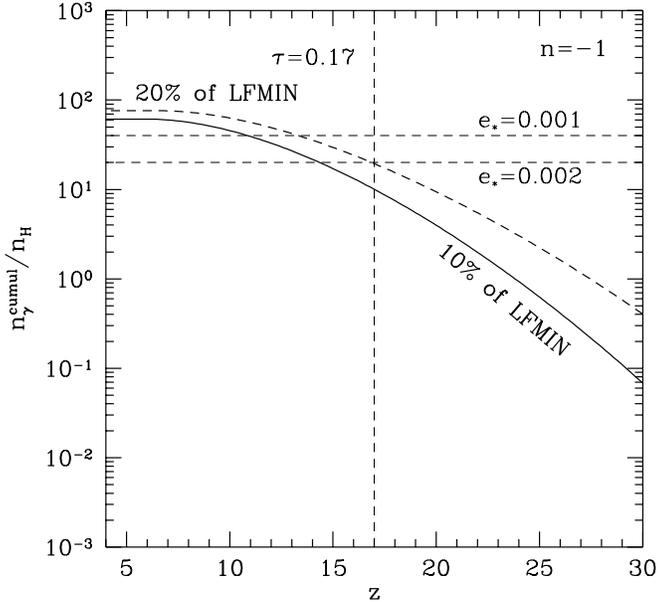,width=90mm}}
\caption{Cumulative photons emitted from Pop III stars with smaller
 normalisation 
for the $n=-1$ isocurvature model. Only $1\%$ and $2\%$ of Seljak's analysis of Ly$\alpha$
forest are needed for $10$ and $20$ photons par H atom at $z=17$.
}
\label{fig4}
\end{figure}
We can clearly see the flattening of number of photons at $z \simgt 7$.
This flattening may allow the small neutral H fraction which is observed
in the spectra of the highest redshift SDSS QSOs, via the Gunn-Peterson
effect.

\subsection{Halo microlensing}

The flux ratios of several quadruple-lensed quasars can only be interpreted
if halo substructure is adding differentially to the lensing optical
depth. Between 0.6 and 7 percent of the halo mass is required to be in
structures of mass up to $10^8 -10^{10} \rm M_\odot,$ within a projected
radius of 10 kpc of a massive halo at $z=0.6$~cite{dal,met}.  Most of the
contribution to the optical depth comes from within the scale radius of the
dark halo, since at larger radii the mean halo density decreases as $r^{-3}.$
However the numerical simulations do not have the resolution to tell whether
the halo substructure survives, for a canonical scale-invariant initial
spectrum of fluctuations. Semi-analytical methods suggest that the
substructure fraction is insensitive to tilt or roll, but possibly too low
(Zentner and Bullock 2003)  for the purely adiabatic model.

The model advocated  here can readily accommodate the needs of
halo substructure lensing, as the early forming substructures are more 
numerous and denser, and so resistant to tidal disruption. 

\subsection{The mass fraction in minihalos, Populations II and III at high
$z$}

We may define minihalos to be dark matter clouds which are below the mass
threshold for star formation. The relevant mass range for minihalos that can
trap baryons requires temperatures above that of the CMB and 
 masses above about $10^4 h^{-1}\rm M_\odot$.
In contrast, cooling is only effective at masses above approximately $10^6 \rm M_\odot.$

The abundance of minihalos is shown in Figure 5 as a function of redshift
in a typical isocurvature/adiabatic model.
Note that they are more numerous out to $z\sim 40$ than the peak in the $\Lambda$CDM
model, which occurs for minihalos at $z\sim 10.$
\begin{figure}
\centerline{\psfig{file=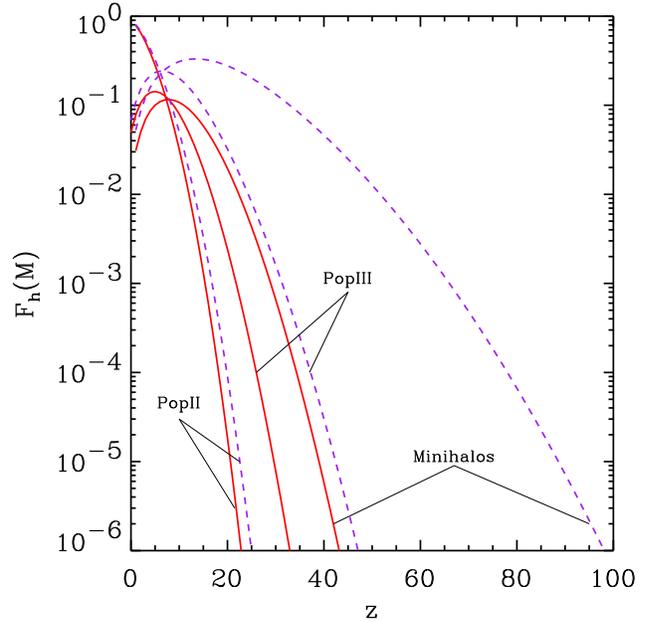,width=90mm}}
\caption{
The mass fraction of halos,
within a given mass range, is shown as a function of redshift for the
PL$\Lambda$CDM model (solid lines) and the $n=-1.5$ isocurvature
reionisation model (dashes lines). For each model 3 cases are
presented: mass intervals defined by all masses with $T_{vir} > 10^4
K$ (predominantly halos that cool via Ly$\alpha$ and form popII
stars); masses with $T_{vir} < 10^4 K$ and above $10^6 h^{-1}\rm M_{\odot}$
(predominantly halos that cool via $\rm H_2$ and form popIII stars); and
minihalos of mass above $10^4 h^{-1}\rm M_{\odot}$ and less than $10^6 h^{-1} 
\rm M_{\odot}$
(halos that are too low in mass to cool) but may contain residual
trapped gas.
}
\label{fig5}
\end{figure}

Also shown in Figure 5 are the mass fractions in Population III and in
Population II stars that form in dwarf galaxy halos. These are defined
by the respective criteria that cooling by $\rm H_2$ and Ly$\alpha$
cooling are the dominant dissipative mechanisms for concentrating the
baryons and enabling fragmentation to proceed.  We base our criteria for
formation of Population III and Population II stars in primordial clouds
on the formulation by Haiman and Holder (2003) in terms of Type II vs
Type I halos. Their classification is based on the distinction between
$\rm H_2$ and $\rm HI$ cooling: we simply take this definition to its
logical conclusion, given that the consensus view is that molecular
cooling results in very massive stars (Pop III) and atomic cooling
allows fragmentation to the ``normal'' mass
range~\cite{abel,bromm2,omu}.  The corresponding mass ranges are defined
by all masses with $T_{vir} < 10^4 K$ and above $10^6 h^{-1} \rm M_{\odot}$
(predominantly halos that cool via $\rm H_2$ and form popIII stars), and
by all masses with $T_{vir} > 10^4 K$ (predominantly halos that cool via
Ly$\alpha$ and form Population II stars).

We see that  Population III  stars are boosted by an order of magnitude at
$z\sim 20$, although Population II star formation is not greatly affected
by the isocurvature admixture relative to $\Lambda$CDM. This is because the
mass fraction in the relatively massive clouds required in this latter case,
typically  in excess of 
$\sim 10^9 h^{-1} \rm M_{\odot}$, 
is strongly constrained by our model which incorporates the requirement that
we cannot overly perturb 
the Ly$\alpha$ forest.

\subsection{Baryon trapping and SZ fluctuations}

We consider the effects of baryon trapping in the isocurvature
 perturbation-induced substructure.  This will have the effect of enhancing
the temperature and SZ fluctuations produced at reionisation relative to
those predicted for the pure adiabatic case.  Even if the baryons cannot
cool, they are trapped at high redshift in dark matter minihalos of mass
above about $10^4 h^{-1}\rm M_\odot.$ The baryon overdensity is $\sim
(\sigma_v/v_s)^2,$ where $\sigma_v$ is the velocity dispersion in the dark
matter minihalo and $v_s$ is the gas sound velocity. Trapping occurs only if
$T >T_{\rm CMB},$ and this happens in the more massive minihalos
e.g. above $10^4 h^{-1}\rm M_\odot$.  Cooling via $\rm H_2$
further enhances the gas density in
late-forming minihalos.
It has been argued that Population III stars form in such gas, whereas
once the $\rm H_2$ is photo-dissociated , $\rm H$ cooling predominates  via
Ly$\alpha$ excitations and
stars of lower mass can form as fragmentation continues to higher
densities~\cite{omu}.

However star formation is by no means guaranteed. Minihalos may retain
gas supported in a stable configuration by  dark matter
 self-gravity~\cite{ume,ger}.In the present case, such minihalos
 could be very abundant at $z>20,$ and may provide a unique window on
 the dark ages of the early universe via radio and NIR observations
of a diffuse background of  redshifted 21cm and Ly$\alpha$ emission.
For example with 100  Ly$\alpha$ photons per baryon one might see at
$2\mu$ a 1 percent contribution to the diffuse extragalactic background, which
amounts to $\nu i_\nu \approx 10 \rm nw m^{-2} sr^{-1}$, but  
could however  be spectrally concentrated in a feature with width $\Delta
\nu/\nu\sim 0.1$ associated with the epoch of reionisation.


\section*{Acknowledgments}

We thank R. Somerville for useful discussions.  NS is supported by
Japanese Grant-in-Aid for Science Research Fund of the Ministry of
Education, No.14340290.  NS also thanks Max Planck Institute for
Astrophysics and Astrophysics Department of Oxford University for their
kind hospitality.  SZ acknowledges hospitality of the National 
Astronomical Observatory Japan.  
JS thanks JSPS for his visit to the National Astronomical
Observatory Japan where this paper was conceived.

\end{document}